\documentclass{article}

     \usepackage[final]{neurips_2019}

\usepackage[utf8]{inputenc} 
\usepackage[T1]{fontenc}    
\usepackage{hyperref}       
\usepackage{url}            
\usepackage{booktabs}       
\usepackage{amsfonts}       
\usepackage{nicefrac}       
\usepackage{microtype}      
\usepackage{graphicx}

\usepackage{bm} 
\usepackage{amssymb}
\usepackage{amsthm}
\usepackage{amsmath}

\title{DP-LSTM: Differential Privacy-inspired LSTM for Stock Prediction Using Financial News}

\author{%
  Xinyi Li$^1$, Yinchuan Li$^{2,1}$, Hongyang Yang$^1$, Liuqing Yang$^1$, Xiao-Yang Liu$^1$ \\
  $^1$Columbia University, $^2$Beijing Institute of Technology\\
  \texttt{\{xl2717, yl3923, hy2500, ly2335, xl2427\}@columbia.edu}
  }

\begin{document}

\maketitle

\begin{abstract}
Stock price prediction is important for value investments in the stock market. In particular, short-term prediction that exploits financial news articles is promising in recent years. In this paper, we propose a novel deep neural network DP-LSTM for stock price prediction, which incorporates the news articles as hidden information and integrates difference news sources through the differential privacy mechanism. First, based on the autoregressive moving average model (ARMA), a sentiment-ARMA is formulated by taking into consideration the information of financial news articles in the model. Then, an LSTM-based deep neural network is designed, which consists of three components: LSTM, VADER model and differential privacy (DP) mechanism. The proposed DP-LSTM scheme can reduce prediction errors and increase the robustness. Extensive experiments on S\&P 500 stocks show that (i) the proposed DP-LSTM achieves 0.32\% improvement in mean MPA of prediction result, and (ii) for the prediction of the market index S\&P 500, we achieve up to 65.79\% improvement in MSE.
\end{abstract}

\section{Introduction}

Stock prediction is crucial for quantitative analysts and investment companies. Stocks' trends, however, are affected by a lot of factors such as interest rates, inflation rates and financial news [12]. To predict stock prices accurately, one must use these variable information. In particular, in the banking industry and financial services, analysts' armies are dedicated to pouring over, analyzing, and attempting to quantify qualitative data from news. A large amount of stock trend information is extracted from the large amount of text and quantitative information that is involved in the analysis.

Investors may judge on the basis of technical analysis, such as charts of a company, market indices, and on textual information such as news blogs or newspapers. It is however difficult for investors to analyze and predict market trends according to all of these information [22]. A lot of artificial intelligence approaches have been investigated to automatically predict those trends [3]. For instance, investment simulation analysis with artificial markets or stock trend analysis with lexical cohesion based metric of financial news' sentiment polarity. Quantitative analysis today is heavily dependent on data. However, the majority of such data is unstructured text that comes from sources like financial news articles. The challenge is not only the amount of data that are involved, but also the kind of language that is used in them to express sentiments, which means emoticons. Sifting through huge volumes of this text data is difficult as well as time-consuming. It also requires a great deal of resources and expertise to analyze all of that [4].

To solve the above problem, in this paper we use sentiment analysis to extract information from textual information. Sentiment analysis is the automated process of understanding an opinion about a given subject from news articles [5]. The analyzed data quantifies reactions or sentiments of the general public toward people, ideas or certain products and reveal the information's contextual polarity. Sentiment analysis allows us to understand if newspapers are talking positively or negatively about the financial market, get key insights about the stock's future trend market.

We use valence aware dictionary and sentiment reasoner (VADER) to extract sentiment scores. VADER is a lexicon and rule-based sentiment analysis tool attuned to sentiments that are expressed in social media specifically [6]. VADER has been found to be quite successful when dealing with NY Times editorials and social media texts. This is because VADER not only tells about the negativity score and positively but also tells us about how positive or negative a sentiment is.

However, news reports are not all objective. We may increase bias because of some non-objective reports, if we rely on the information that is extracted from the news for prediction fully. Therefore, in order to enhance the prediction model's robustness, we will adopt differential privacy (DP) method. DP is a system for sharing information about a dataset publicly by describing groups' patterns within the dataset while withholding information about individuals in the dataset. DP can be achieved if the we are willing to add random noise to the result. For example, rather than simply reporting the sum, we can inject noise from a Laplace or gaussian distribution, producing a result that’s not quite exact, that masks the contents of any given row.

In the last several years a promising approach to private data analysis has emerged, based on DP, which ensures that an analysis outcome is "roughly as likely" to occur independent of whether any individual opts in to, or to opts out of, the database. In consequence, any one individual's specific data can never greatly affect the results. General techniques for ensuring DP have now been proposed, and a lot of datamining tasks can be carried out in a DP method, frequently with very accurate results [21]. We proposed a DP-LSTM neural network, which increase the accuracy of prediction and robustness of model at the same time.

The remainder of the paper is organized as follows. In Section 2, we introduce stock price model, the sentiment analysis and differential privacy method. In Section 3, we develop the different privacy-inspired LSTM (DP-LSTM) deep neural network and present the training details. Prediction results are provided in Section 4. Section 5 concludes the paper.

\section{Problem Statement}
\label{gen_inst}

In this section, we first introduce the background of the stock price model, which is based on the autoregressive moving average (ARMA) model. Then, we present the sentiment analysis details of the financial news and introduce how to use them to improve prediction performance. At last, we introduce the differential privacy framework and the loss function.

\subsection{ARMA Model}
The ARMA model, which is one of the most widely used linear models in time series prediction [17], where the future value is assumed as a linear combination of the past errors and past values. ARMA is used to set the stock midterm prediction problem up. Let ${X}_t^\text{A}$ be the variable based on ARMA at time $t$, then we have
\begin{align}
\label{ARIMA}
    {X}_t^\text{A}=&~ f_1(\{X_{t-i}\}_{i=1}^p) = \mu + \sum_{i=1}^{p}\phi_i X_{t-i} - \sum_{i=1}^{q}\psi_j\epsilon_{t-j}+\epsilon_t,
\end{align}
where $X_{t-i}$ denotes the past value at time $t-i$; $\epsilon_{t}$ denotes the random error at time $t$; $\phi_i$ and $\psi_j$ are the coefficients; $\mu$ is a constant; $p$ and $q$ are integers that are often referred to as autoregressive and moving average polynomials, respectively.

\begin{figure}
  \centering
  \includegraphics[width=\linewidth]{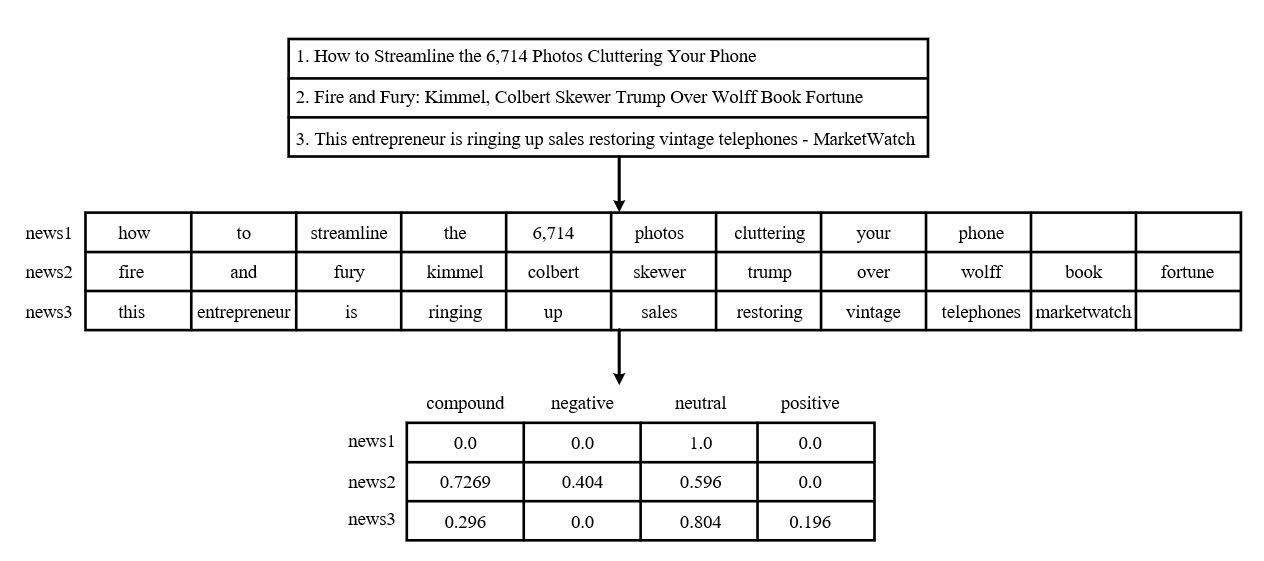}
  \caption{NLTK processing. For preprocessing, each news title will be tokenized into individual words. Then applying SentimentIntensityAnalyzer from NLTK vadar to calculate the polarity score.}
\label{figure:NLTK}
\end{figure}

\subsection{Sentiment Analysis}

Another variable highly related to stock price is the textual information from news, whose changes may be a precursor to price changes. 
In our paper, news refers to a news article's title on a given trading day. It has been used to infer whether an event had informational content and whether investors' interpretations of the information were positive, negative or neutral. We hence use sentiment analysis to identify and extract opinions within a given text. Sentiment analysis aims at gauging the attitude, sentiments, evaluations and emotions of a speaker or writer based on subjectivity's computational treatment in a text [19]-[20].

\begin{figure}
  \centering
  \includegraphics[width=0.40\linewidth]{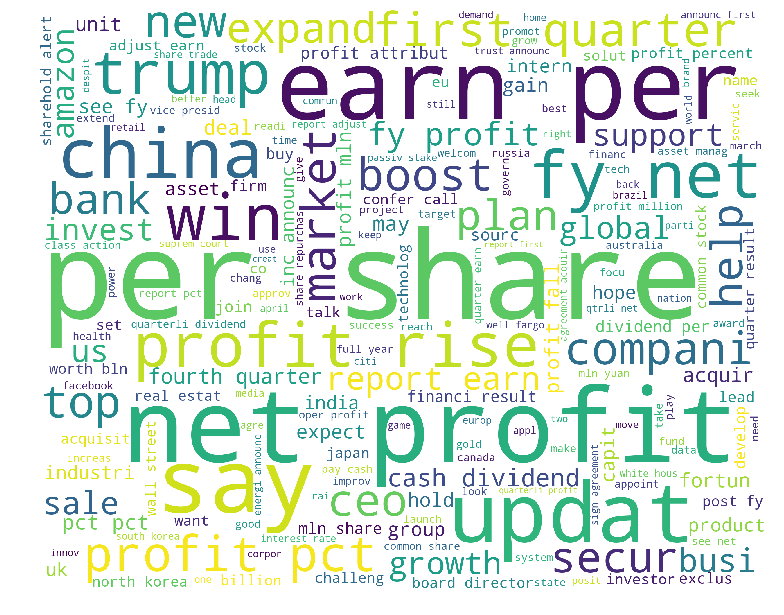}
     \includegraphics[width=0.40\linewidth]{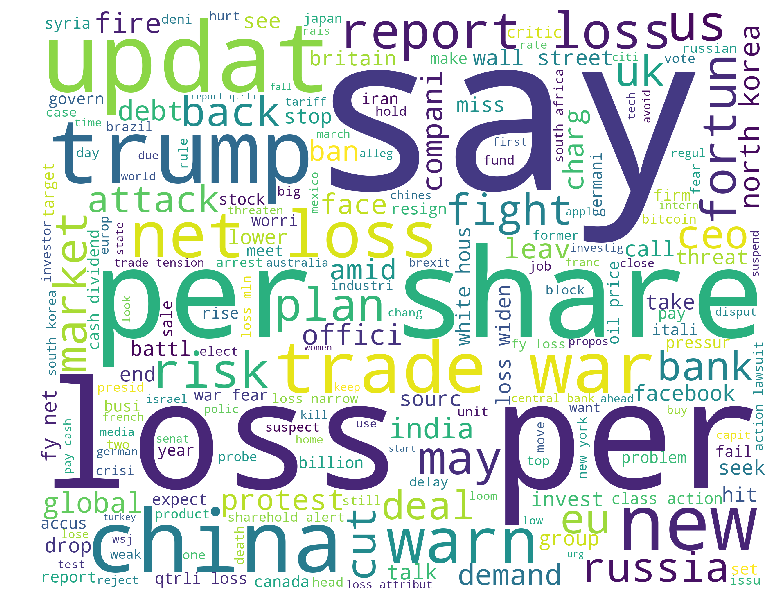}
  \caption{Positive wordcloud (left) and negative wordcloud (right).
  We divide the news based on their compound score. For both positive news and negative news, we count all the words and rank them to create the wordcloud. The larger the word, more frequently it has appeared in the source. 
  }
\label{figure:cloud}
\end{figure}

Figure~\ref{figure:NLTK} shows an example of the sentiment analysis results obtained from financial news titles that were based on VADER. VADER uses a combination of a sentiment lexicon which are generally labelled according to their semantic orientation as either negative or positive.
VADER has been found to be quite successful when dealing with news reviews. It is fully open-sourced under the MIT License. The result of VADER represent as sentiment scores, which include the positive, negative and neutral scores represent the proportion of text that falls in these categories. This means all these three scores should add up to 1. Besides, the Compound score is a metric that calculates the sum of all the lexicon ratings which have been normalized between -1(most extreme negative) and +1 (most extreme positive). Figure~\ref{figure:cloud} shows the positive and negative wordcloud, which is an intuitive analysis of the number of words in the news titles.

\subsection{Sentiment-ARMA Model and Loss Function}

To take the sentiment analysis results of the financial news into account, we introduce the sentiment-ARMA model as follows
\begin{align}
\label{ARIMA_SPY_V}
	\hat{X}_t  = \alpha {X}_t^{\text{A}} + \lambda {S}_t^{\text{A}}  + c = \alpha {X}_t^{\text{A}} + \lambda f_2(\underbrace{ \{S_{t-i}\}_{i=1}^p }_{\text{Sentiment}}) + c,
\end{align}
where $\alpha$ and $\lambda$ are weighting factors; $c$ is a constant; and $f_2(\cdot)$ is similar to $f_1(\cdot)$ in the ARMA model \eqref{ARIMA} and is used to describe the prediction problem.

In this paper, the LSTM neural network is used to predict the stock price, the input data is the previous stock price and the sentiment analysis results. Hence, the sentiment based LSTM neural network (named sentiment-LSTM) is aimed to minimize the following loss function:
\begin{align}
\label{L2}
    {\cal L} =&~ \min\sum_{t=p+1}^{p+T} \left\| X_t- \hat X_t \right\|_2^2,
\end{align}
where $T$ denotes the number of prediction time slots, i.e., $t = 1,...,p$ are the observations (training input data), $t = p+1,...,p+T$ are the predicts (training output data); and $\hat X_t$ is given in \eqref{ARIMA_SPY_V}.

\subsection{Overview of LSTM}

\begin{figure}
  \centering
  \includegraphics[width=3.0in]{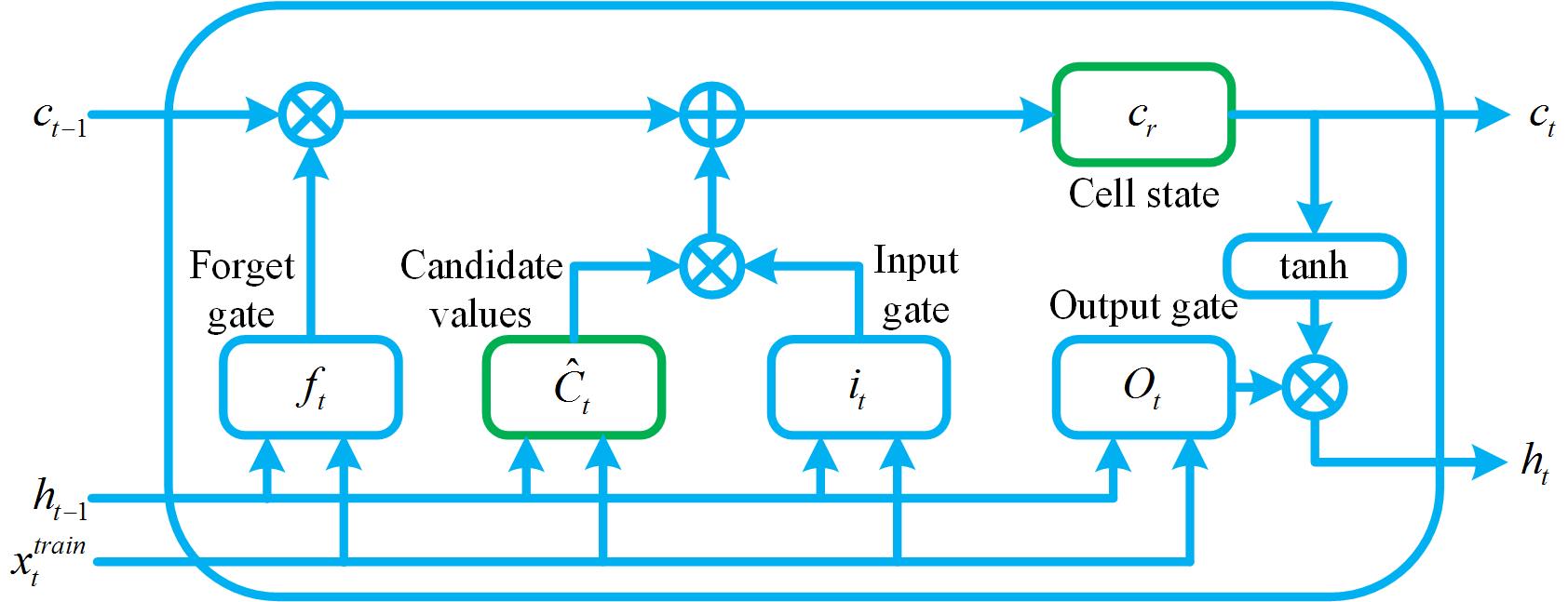}
  \caption{LSTM procedure}
  \label{figure:Mid-LSTM}
\end{figure}

Denote $\mathcal{X}_t^{\text{train}} = \{X_{t-i},S_{t-i}\}_{i=1}^p$ as the training input data. Figure~\ref{figure:Mid-LSTM} shows the LSTM's structure network, which comprises one or more hidden layers, an output layer and an input layer [16]. LSTM networks' main advantage is that the hidden layer comprises memory cells. Each memory cell recurrently has a core self-connected linear unit called `` Constant Error Carousel (CEC)'' [13], which provides short-term memory storage and has three gates:

\begin{itemize}

\item Input gate, which controls the information from a new input to the memory cell, is given by
\begin{align}
\label{LSTM1_1}
   i_t =&~ \sigma(W_i\times[h_{t-1},{\cal X}_t^{\text{train}}]+b_i), \\
\label{LSTM1_2}
   \hat{c}_t = &~\tanh(W_c\times[h_{t-1},{\cal X}_t^{\text{train}}]+b_c),
\end{align}
where $h_{t-1}$ is the hidden state at the time step $t-1$; $i_t$ is the output of the input gate layer at the time step $t$; $\hat{c}_t$ is the candidate value to be added to the output at the time step $t$; $b_i$ and $b_c$ are biases of the input gate layer and the candidate value computation, respectively; $W_i$ and $W_c$ are weights of the input gate and the candidate value computation, respectively; and $\sigma(x) = 1/(1+e^{-x})$ is the pointwise nonlinear activation function.

\item Forget gate, which controls the limit up to which a value is saved in the memory, is given by
\begin{align}
\label{LSTM2_1}
   f_t = \sigma(W_f\times[h_{t-1},{\cal X}_t^{\text{train}}]+b_f),
\end{align}
where $f_t$ is the forget state at the time step $t$, $W_f$ is the weight of the forget gate; and $b_f$ is the bias of the forget gate.

\item Output gate, which controls the information output from the memory cell, is given by
\begin{align}
\label{LSTM3_1}
   c_t =&~ f_t \times c_{t-1}+i_t \times \hat{c}_t,\\
   o_t =&~ \sigma(W_o\times[h_{t-1},{\cal X}_t^{\text{train}}]+b_o),\\
   h_t =&~ o_t \times \tanh(c_t),
\label{LSTM3_2}
\end{align}
where new cell states $c_t$ are calculated based on the results of the previous two steps; $o_t$ is the output at the time step $t$; $W_o$ is the weight of the output gate; and $b_o$ is the bias of the output gate [14].

\end{itemize}

\subsection{Definition of Differential Privacy}
Differential privacy is one of privacy's most popular definitions today, which is a system for publicly sharing information about a dataset by describing the patterns of groups within the dataset while withholding information about individuals in the dataset. It intuitively requires that the mechanism that outputs information about an underlying dataset is robust to one sample's any change, thus protecting privacy. A mechanism ${f}$ is a random function that takes a dataset $\mathcal{N}$ as input, and outputs a random variable ${f}(\mathcal{N})$. For example, suppose $\mathcal{N}$ is a news articles dataset, then the function that outputs compound score of articles in $\mathcal{N}$ plus noise from the standard normal distribution is a mechanism [7].

Although differential privacy was originally developed to facilitate secure analysis over sensitive data, it can also enhance the robustness of the data. Note that finance data, especially news data and stock data, is unstable with a lot of noise, with a more robust data the accuracy of prediction will be improved. Since we predict stock price by fusing news come from different sources, which might include fake news. Involving differential privacy in the training to improve the robustness of the finance news is meaningful.

\section{Training DP-LSTM Neural Network}
\label{headings}

It is known that it is risky to predict stocks by considering news factors, because news can't guarantee full notarization and objectivity, many times extreme news will have a big impact on prediction models. To solve this problem, we consider entering the idea of the differential privacy when training. In this section, our DP-LSTM deep neural network training strategy is presented. The input data consists of three components: stock price, sentiment analysis compound score and noise.

\subsection{Data Preprocessing and Normalization}

\subsubsection{Data Preprocessing}

The data for this project are two parts, the first part is the historical S\&P 500 component stocks, which are downloaded from the Yahoo Finance. We use the data over the period of from 12/07/2017 to 06/01/2018. The second part is the news article from financial domain are collected with the same time period as stock data. Since our paper illustrates the relationship between the sentiment of the news articles and stocks' price. Hence,  only news article from financial domain are collected. 
The data is mainly taken from Webhose archived data, which consists of 306242 news articles present in JSON format, dating from December 2017 up to end of June 2018. The former 85\% of the dataset is used as the training data and the remainder 15\% is used as the testing data.
The News publishers for this data are CNBC.com, Reuters.com, WSJ.com, Fortune.com. The Wall Street Journal is one of the largest newspapers in the United States, which coverage of breaking news and current headlines from the US and around the world include top stories, photos, videos, detailed analysis and in-depth thoughts; CNBC primarily carries business day coverage of U.S. and international financial markets, which following the end of the business day and on non-trading days; Fortune is an American multinational business magazine; Reuters is an international news organization. We preprocess the raw article body and use NLTK sentiment package alence Aware Dictionary and Sentiment Reasoner (VADER) to extract sentiment scores.

 The stocks with missing data are deleted, and the dataset we used eventually contains 451 stocks and 4 news resources (CNBC.com, Reuters.com, WSJ.comFortune.com.). Each stock records the adjust close price and news compound scores of 121 trading days.

\begin{figure}[!h]
	\centering
	\includegraphics[width=3.0in]{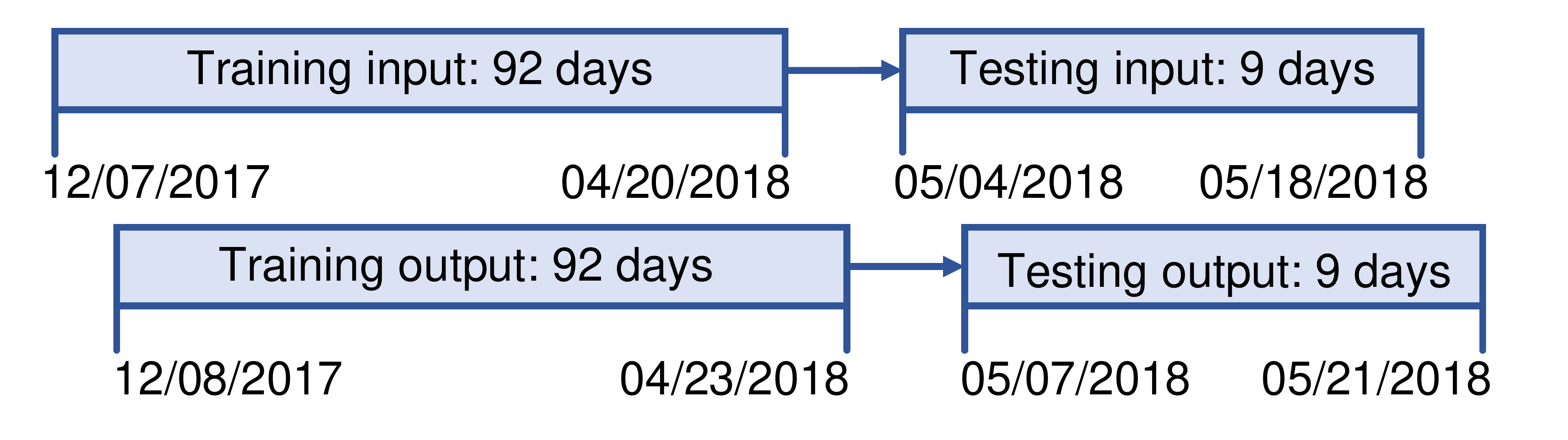}
	\caption{Schematic diagram of rolling window.}
	\label{figure:rolling}
\end{figure}

A rolling window with size $10$ is used to separate data, that is, We predict the stock price of the next trading day based on historical data from the previous 10 days, hence resulting in a point-by-point prediction~[15]. In particular, the training window is initialized with all real training data. Then we shift the window and add the next real point to the last point of training window to predict the next point and so forth. Then, according to the length of the window, the training data is divided into 92 sets of training input data (each set length 10) and training output data (each set length 1). The testing data is divided into input and output data of 9 windows (see Figure~\ref{figure:rolling}).

\subsubsection{Normalization}

To detect stock price pattern, it is necessary to normalize the stock price data. Since the LSTM neural network requires the stock patterns during training, we use ``min-max'' normalization method to reform dataset, which keeps the pattern of the data [11], as follow:
\begin{align}
\label{N3}
 X_{t}^{n}=\frac{X_{t}-\min({X_{t}})}{\max({X_{t}})-\min({X_{t}})},
\end{align}
where $X_{t}^{n}$ denotes the data after normalization. Accordingly, de-normalization is required at the end of the prediction process to get the original price, which is given by
\begin{align}
\label{N4}
 \hat{X}_{t}=\hat{X}_{t}^{n}[\max({X}_{t})-\min({X}_{t})]+\min({X}_{t}),
\end{align}
where $\hat{X}_{t}^{n}$ denotes the predicted data and $\hat{X}_{t}$ denotes the predicted data after de-normalization.

Note that compound score is not normalized, since the compound score range from -1 to 1, which means all the compound score data has the same scale, so it is not require the normalization processing.

\subsection{Adding Noise}
We consider the differential privacy as a method to improve the robustness of the LSTM predictions~[8]. We explore the interplay between machine learning and differential privacy, and found that differential privacy has several properties that make it particularly useful in application such as robustness to extract textual information [9]. The robustness of textual information means that accuracy is guaranteed to be unaffected by certain false information [10].

The input data of the model has 5 dimensions, which are the stock price and four compound scores as $(X^t, S_1^t, S_2^t, S_3^t, S_4^t), t=1,...,T$, where $X^t$ represents the stock price and $S_i^t,~i=1,...,4$ respectively denote the mean compound score calculated from WSJ, CNBC, Fortune and Reuters. According to the process of differential privacy, we add Gaussian noise with different variances to the news according to the variance of the news, i.e., the news compound score after adding noise is given by
\begin{align}
\label{N4}
 \widetilde S_i^t= S_i^t + \mathcal{N}(0,\lambda \text{var}(S_i)),~i=1,...,4,
\end{align}
where $\text{var}(\cdot)$ is the variance operator, $\lambda$ is a weighting factor and $\mathcal{N}(\cdot)$ denotes the random Gaussian process with zero mean and variance $\lambda \text{var}(S_i)$.

We used python to crawl the news from the four sources of each trading day, perform sentiment analysis on the title of the news, and get the compound score. After splitting the data into training sets and test sets, we separately add noise to each of four news sources of the training set, then, for $n$-th stock, four sets of noise-added data $(X^n_t, {\widetilde S^t_1}, S^t_2, S^t_3, S^t_4)$, $(X^n_t, {S^t_1}, \widetilde S^t_2, S^t_3, S^t_4)$, $(X^n_t, { S^t_1}, S^t_2, \widetilde S^t_3, S^t_4)$, $(X^n_t, { S^t_1}, S^t_2, S^t_3, \widetilde S^t_4)$ are combined into a new training data through a rolling window. The stock price is then combined with the new compound score training data as input data for our DP-LSTM neural network.

\subsection{Training Setting}
The LSTM model in figure~\ref{figure:Mid-LSTM} has six layers, followed by an LSTM layer, a dropout layer, an LSTM layer, an LSTM layer, a dropout layer, a dense layer, respectively. The dropout layers (with dropout rate 0.2) prevent the network from overfitting. The dense layer is used to reshape the output. Since a network will be difficult to train if it contains a large number of LSTM layers~[16], we use three LSTM layers here.

In each LSTM layer, the loss function is the mean square error (MSE), which is the sum of the squared distances between our target variable and the predicted value. In addition, the ADAM~[17] is used as optimizer, since it is straightforward to implement, computationally efficient and well suited for problems with large data set and parameters.

There are many methods and algorithms to implement sentiment analysis systems. In this paper, we use rule-based systems that perform sentiment analysis based on a set of manually crafted rules. Usually, rule-based approaches define a set of rules in some kind of scripting language that identify subjectivity, polarity, or the subject of an opinion. We use VADER, a simple rule-based model for general
sentiment analysis.

\begin{figure}
  \centering
  \includegraphics[width=3.0in]{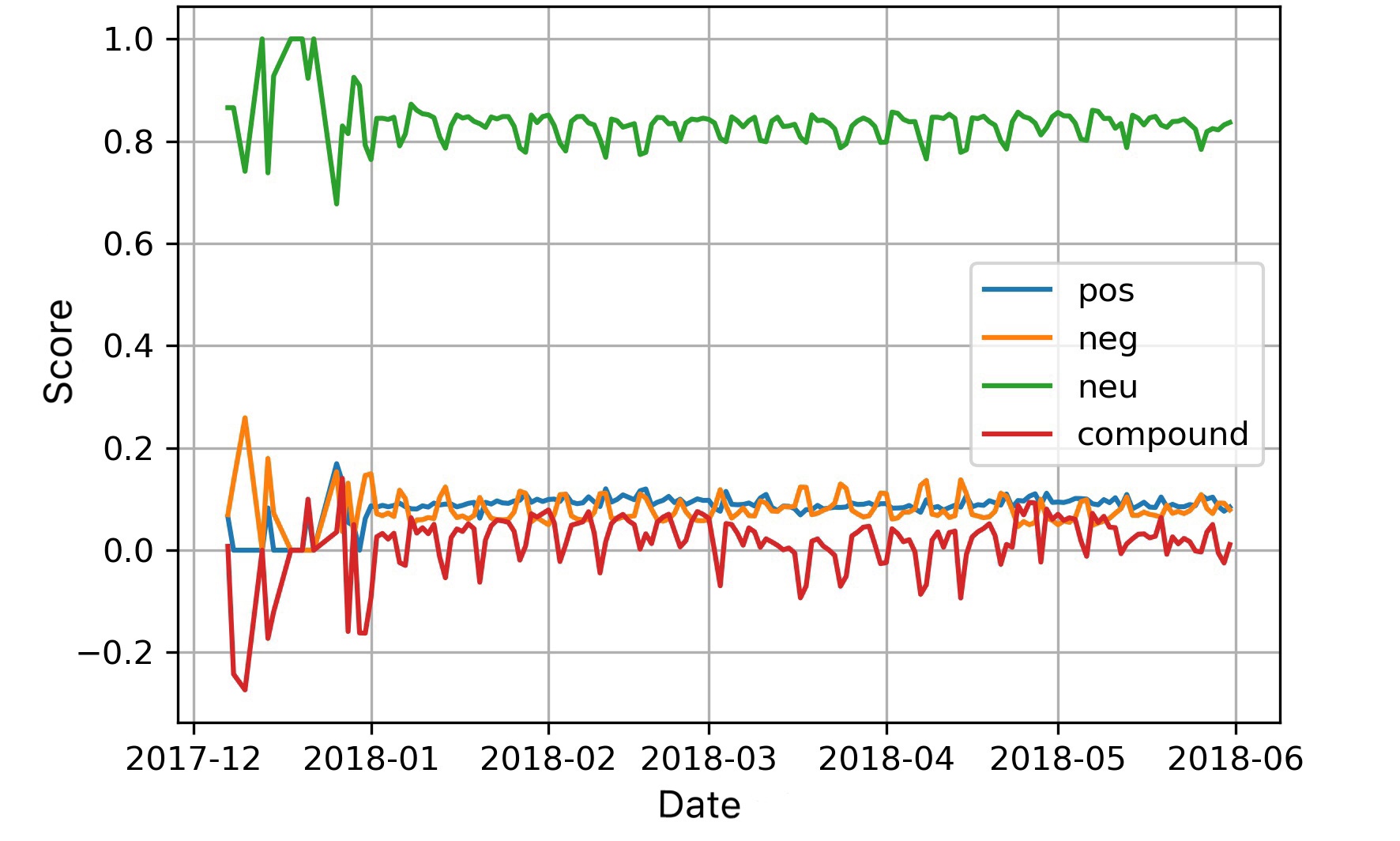}
  \caption{NLTK result.}
  \label{nltk}
\end{figure}

\section{Performance Evaluation}
\label{others}

In this section, we validate our DP-LSTM based on the S\&P 500 stocks. 
We calculate the mean prediction accuracy (MPA) to evaluate the proposed methods, which is defined as 
\begin{align}
\label{accuracy1}
\text{MPA}_{t} = 1 - \frac{1}{L}\sum_{\ell=1}^{L} \frac{|X_{t,\ell}-\hat{X}_{t,\ell}|}{X_{t,\ell}},
\end{align}
where $X_{t,\ell}$ is the real stock price of the $\ell$-th stock on the $t$-th day, $L$ is the number of stocks and $\hat{X}_{t,\ell}$ is the corresponding prediction result. 

Figure~\ref{nltk} plots the average score for all news on the same day over the period. The compound score is fluctuating between -0.3 and 0.15, indicating an overall neutral to slightly negative sentiment. The Positive, Negative and Neutral scores represent the proportion of text that falls in these categories. 
The Compound score is a metric that calculates the sum of all the lexicon ratings which have been normalized between -1 (most extreme negative) and +1 (most extreme positive).

\begin{figure*}[!htb]
	\centering
	\includegraphics[width=3.5in]{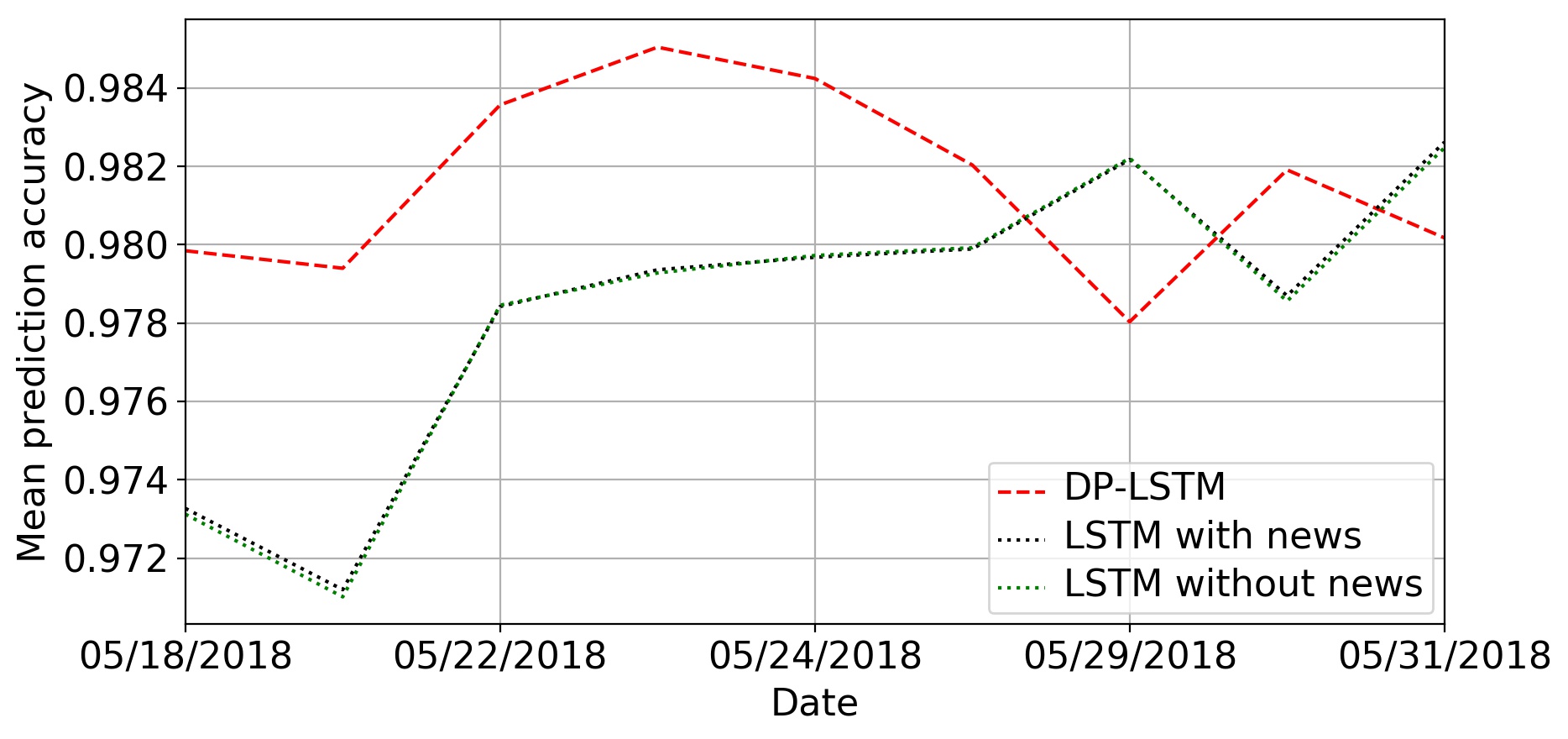}
	\caption{Mean prediction accuracies of the DP-LSTM and vanilla LSTM.}
	\label{figure:MPA-results}
\end{figure*}

Figure~\ref{figure:MPA-results} shows the $\text{MPAs}$ of the proposed DP-LSTM and vanilla LSTM for comparison.  In Table~\ref{tab:MPA}, we give the mean MPA results for the prediction prices, which shows the accuracy performance of DP-LSTM is 0.32\% higer than the LSTM with news. The result means the DP framework can make the prediction result more accuracy and robustness.

\begin{table}
\centering
\begin{tabular}{ccc}  
\toprule
Method  & Mean MPA \\
\midrule
LSTM without news     &  0.978305309 \\
LSTM with news     &  0.978366682 \\
\textbf{DP-LSTM}    & \textbf{0.981582666}  \\
\bottomrule
\end{tabular}
\caption{Predicted Mean MPA results.}
\label{tab:MPA}
\end{table}

Note that the results are obtained by running many trials, since we train stocks separately and predict each price individually due to the different patterns and scales of stock prices. This in total adds up to 451 runs. The results shown in Table~\ref{tab:MPA} is the average of these 451 runs. Furthermore, we provide results for 9 duration over a period in Figure~\ref{figure:MPA-results}. The performance of our DP-LSTM is always better than the LSTM with news. Based on the sentiment-ARMA model and adding noise for training, the proposed DP-LSTM is more robust. The investment risk based on this prediction results is reduced.


\begin{figure*}[!htb]
	\centering
	{\includegraphics[width=5.5 in]{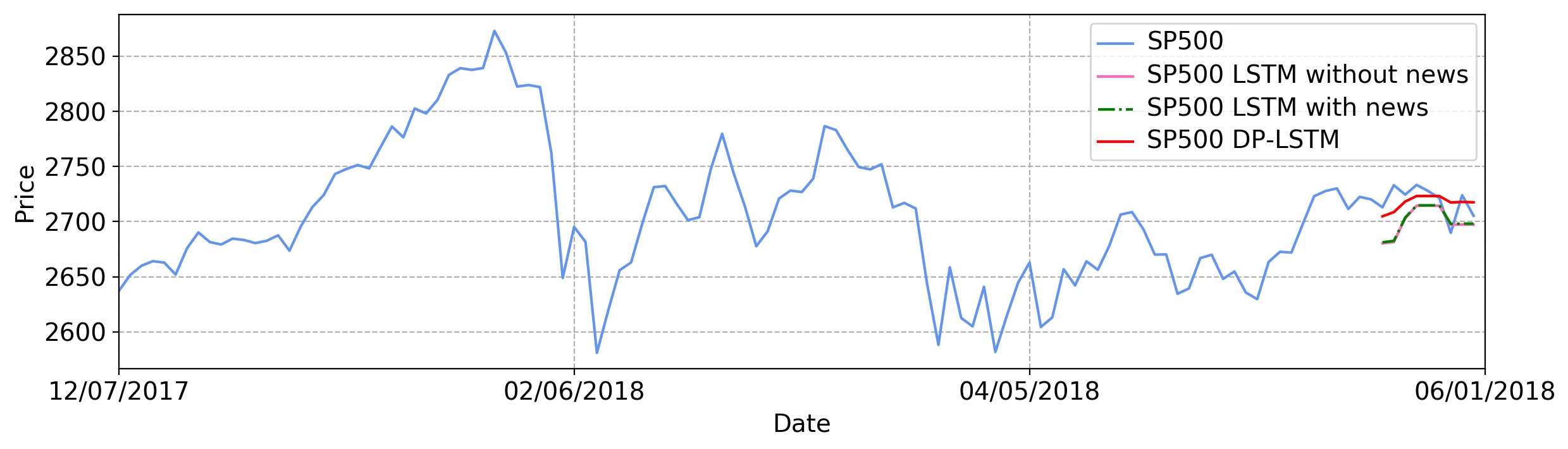}}
	\caption{Prediction result of LSTM based on price.}
	\label{figure:sp500}
\end{figure*}

\begin{table}
\centering
\begin{tabular}{cccc}  
\toprule
Metrics  & LSTM without news & LSTM with news & \textbf{DP-LSTM}\\
\midrule
MSE    & 580.9226827 & 536.6306251 & \textbf{198.7500672}  \\
Accuracy     &  0.99263803 & 0.99292492 & \textbf{0.99582651} \\
Mean error percent     &  0.00736197 & 0.00707508 & \textbf{0.00417349} \\
\bottomrule
\end{tabular}
\caption{S\&P 500 predicted results.}
\label{tab:sp}
\end{table}

In Figure~\ref{figure:sp500}, we can see the prediction results of DP-LSTM with is closer to the real S\&P 500 index price line than other methods. The two lines (prediction results of LSTM with news and LSTM without news) almost coincide in Figure~\ref{figure:sp500}. We can tell the subtle differences from the Table~\ref{tab:sp}, that DP-LSTM is far ahead, and LSTM with news is slightly better than LSTM without news.

\section{Conclusion}

In this paper, we integrated the deep neural network with the famous NLP models (VADER) to identify and extract opinions within a given text, combining the stock adjust close price and compound score to reduce the investment risk. We first proposed a sentiment-ARMA model to represent the stock price, which incorporates influential variables (price and news) based on the ARMA model. Then, a DP-LSTM deep neural network was proposed to predict stock price according to the sentiment-ARMA model, which combines the LSTM, compound score of news articles and differential privacy method. News are not all objective. If we rely on the information extracted from the news for prediction fully, we may increase bias because of some non-objective reports. Therefore, the DP-LSTM enhance robustness of the prediction model. Experiment results based on the S\&P 500 stocks show that the proposed DP-LSTM network can predict the stock price accurately with robust performance, especially for S\&P 500 index that reflects the general trend of the market. S\&P 500 prediction results show that the differential privacy method can significantly improve the robustness and accuracy.

\section*{References}

%
\small

[1] X. Li, Y. Li, X.-Y. Liu, D. Wang, ``Risk management via anomaly circumvent: mnemonic deep learning for midterm stock prediction.'' in {\em Proceedings of 2nd KDD Workshop on Anomaly Detection in Finance (Anchorage ’19)}, 2019.

[2] P. Chang, C. Fan, and C. Liu, ``Integrating a piece-wise linear representation method and a neural network model for stock trading points prediction.'' {\em IEEE Transactions on Systems, Man, and Cybernetics, Part C (Applications and Reviews)} 39, 1 (2009), 80–92.

[3] Akita, Ryo, et al. ``Deep learning for stock prediction using numerical and textual information.'' {\em IEEE/ACIS 15th International Conference on Computer and Information Science (ICIS). IEEE}, 2016.

[4] Li, Xiaodong, et al. ``Does summarization help stock prediction? A news impact analysis.'' {\em IEEE Intelligent Systems} 30.3 (2015): 26-34.

[5] Ding, Xiao, et al. ``Deep learning for event-driven stock prediction.'' {\em Twenty-fourth International Joint Conference on Artificial Intelligence}. 2015.

[6] Hutto, Clayton J., and Eric Gilbert. ``Vader: A parsimonious rule-based model for sentiment analysis of social media text.'' {\em Eighth International AAAI Conference on Weblogs and Social Media}, 2014.

[7] Ji, Zhanglong, Zachary C. Lipton, and Charles Elkan. ``Differential privacy and machine learning: a survey and review.'' {\em arXiv preprint arXiv:1412.7584} (2014).

[8] Abadi, Martin, et al. ``Deep learning with differential privacy.'' {\em Proceedings of the 2016 ACM SIGSAC Conference on Computer and Communications Security}, ACM, 2016.

[9] McMahan, H. Brendan, and Galen Andrew. ``A general approach to adding differential privacy to iterative training procedures.'' {\em arXiv preprint arXiv:1812.06210} (2018).

[10] Lecuyer, Mathias, et al. ``Certified robustness to adversarial examples with differential privacy.'' {\em arXiv preprint arXiv:1802.03471} (2018).

[11] Hafezi, Reza, Jamal Shahrabi, and Esmaeil Hadavandi. ``A bat-neural network multi-agent system (BNNMAS) for stock price prediction: Case study of DAX stock price.'' {\em Applied Soft Computing}, 29 (2015): 196-210.

[12] Chang, Pei-Chann, Chin-Yuan Fan, and Chen-Hao Liu. ``Integrating a piecewise linear representation method and a neural network model for stock trading points prediction.'' {\em IEEE Transactions on Systems, Man, and Cybernetics}, Part C (Applications and Reviews) 39.1 (2008): 80-92.

[13] Gers, Felix A., Nicol N. Schraudolph, and Jürgen Schmidhuber. ``Learning precise timing with LSTM recurrent networks.'' {\em Journal of Machine Learning Research} 3.Aug (2002): 115-143.

[14] Qin, Yao, et al. ``A dual-stage attention-based recurrent neural network for time series prediction.'' {\em arXiv preprint arXiv:1704.02971} (2017).

[15] Malhotra, Pankaj, et al. ``Long short term memory networks for anomaly detection in time series.'' {\em Proceedings. Presses universitaires de Louvain}, 2015.

[16] Sak, Haşim, Andrew Senior, and Françoise Beaufays. ``Long short-term memory recurrent neural network architectures for large scale acoustic modeling.'' {\em Fifteenth annual conference of the international speech communication association}, 2014.

[17] Kingma, Diederik P., and Jimmy Ba. ``Adam: A method for stochastic optimization.'' {\em arXiv preprint arXiv:1412.6980} (2014).

[18] Box, George EP, et al. Time series analysis: forecasting and control. {\em John Wiley \& Sons}, 2015.

[19] Pang, Bo, and Lillian Lee. ``Opinion mining and sentiment analysis.'' {\em Foundations and Trends in Information Retrieval} 2.1–2 (2008): 1-135.

[20] Cambria, Erik. ``Affective computing and sentiment analysis.'' {\em IEEE Intelligent Systems} 31.2 (2016): 102-107.

[21] Dwork C, Lei J. Differential privacy and robust statistics//STOC. 2009, 9: 371-380.

[22] X. Li, Y. Li, Y. Zhan, and X.-Y. Liu. ``Optimistic bull or pessimistic bear: adaptive deep reinforcement learning for stock portfolio allocation.'' in {\em Proceedings of the 36th International Conference on Machine Learning}, 2019.


\end{document}